\documentclass[conference]{IEEEtran}
\IEEEoverridecommandlockouts

\usepackage{algorithmic}
\usepackage{graphicx}
\usepackage{textcomp}
\usepackage{xcolor}
\usepackage{amsmath, calc}
\usepackage{tcolorbox}
\usepackage{multirow}
\usepackage{url}
\usepackage{hyperref}
\usepackage{pgfplots}
\pgfplotsset{compat=1.18}
\usepackage{placeins}
\usepackage{fontawesome}
\usepackage{comment}
\usepackage{tabularx}
\usepackage{enumitem}
\usepackage{booktabs}
\usepackage{framed}
\usepackage{fancybox}
\usepackage{makecell}
\usepackage{csquotes}
\usepackage{orcidlink}
\usepackage{multicol}
\usepackage{colortbl}
\usepackage{amssymb}

\def\BibTeX{{\rm B\kern-.05em{\sc i\kern-.025em b}\kern-.08em
    T\kern-.1667em\lower.7ex\hbox{E}\kern-.125emX}}
\begin{document}

\makeatletter
\newcommand{\linebreakand}{%
  \end{@IEEEauthorhalign}
  \hfill\mbox{}\par
  \mbox{}\hfill\begin{@IEEEauthorhalign}
}
\makeatother

\title{From App Features to Explanation Needs: Analyzing Correlations and Predictive Potential}


\author{
    \IEEEauthorblockN{
        Martin Obaidi\orcidlink{0000-0001-9217-3934},
        Kushtrim Qengaj,
        Jakob Droste\orcidlink{0000-0001-8746-6329}, 
        Hannah Deters\orcidlink{0000-0001-9077-7486}, \\
        Marc Herrmann\orcidlink{0000-0002-3951-3300},
        Elisa Schmid\orcidlink{0009-0006-2498-9986},
        Kurt Schneider\orcidlink{0000-0002-7456-8323}
    }
    \IEEEauthorblockA{
        \textit{Leibniz Universität Hannover, Software Engineering Group} \\
        Hannover, Germany \\
        \{martin.obaidi, jakob.droste, hannah.deters\}@inf.uni-hannover.de, \\
        \{marc.herrmann, elisa.schmid, kurt.schneider\}@inf.uni-hannover.de, \\ kushtrim.qengaj@stud.uni-hannover.de
    }
    \and
    \IEEEauthorblockN{
        Jil Klünder\orcidlink{0000-0001-7674-2930}
    }
    \IEEEauthorblockA{
        \textit{University of Applied Sciences} \\
        \textit{FHDW Hannover} \\
        Hannover, Germany \\
        jil.kluender@fhdw.de
    }
}

\maketitle

\begin{abstract}

In today's digitized world, software systems must support users in understanding both how to interact with a system and why certain behaviors occur. This study investigates whether explanation needs, classified from user reviews, can be predicted based on app properties, enabling early consideration during development and large-scale requirements mining. We analyzed a gold standard dataset of 4,495 app reviews enriched with metadata (e.g., app version, ratings, age restriction, in-app purchases). Correlation analyses identified mostly weak associations between app properties and explanation needs, with moderate correlations only for specific features such as app version, number of reviews, and star ratings. Linear regression models showed limited predictive power, with no reliable forecasts across configurations. Validation on a manually labeled dataset of 495 reviews confirmed these findings. Categories such as \textit{Security \& Privacy} and \textit{System Behavior} showed slightly higher predictive potential, while \textit{Interaction} and \textit{User Interface} remained most difficult to predict. Overall, our results highlight that explanation needs are highly context-dependent and cannot be precisely inferred from app metadata alone. Developers and requirements engineers should therefore supplement metadata analysis with direct user feedback to effectively design explainable and user-centered software systems.

\end{abstract}

\begin{IEEEkeywords}
explainability, requirements engineering, data mining, app reviews
\end{IEEEkeywords}

\section{Introduction}
\label{sec:intro}

Modern software systems are becoming ever more complex and opaque, making it difficult for users to understand and trust system behavior—especially in AI-driven or data-rich domains~\cite{adadi2018peeking,Antinyan2020complex,levy2021understanding}. As a result, \textit{explainability} has emerged as a crucial non-functional requirement, supporting transparency, trust, and regulatory needs~\cite{kohl2019explainability,chazette2021exploring,Deters2025quality,deters2024qualitymodel,deters2025identifying}. Explanations enhance not just comprehension but also usability, safety, and satisfaction, notably in sensitive areas like healthcare or privacy~\cite{Biswas2021,brunotte2023privacy,brunotte_quo_2022}.

Despite its importance, explainability remains challenging to operationalize in requirements engineering (RE)~\cite{Frattini22,chazette2022can}. User needs for explanations are often expressed informally, through app reviews~\cite{obaidi2025appKnowledge}, support requests, or other feedback channels~\cite{anders2022userfeedback,anders2023userfeedback,obaidi2025elicit}, making systematic identification and structuring difficult~\cite{unterbusch2023explanation,obaidi2025automatingexplanationneedmanagement}. Needs also vary by context, task, expertise, and even mood~\cite{droste2023designing,ramos2021modeling,xu2019roleOfUserMood}, further complicating requirements elicitation and targeted design.

Crowd-based Requirements Engineering (CrowdRE) leverages user-generated content such as app reviews as a rich source for uncovering implicit explanation needs~\cite{droste2024explanations,Droste2025REJexpl,unterbusch2023explanation}. Taxonomies help structure and automate the identification of these needs, for example by classifying explicit/implicit requests or mapping needs to system aspects like \textit{User Interface}, \textit{Security}, or \textit{Interaction}~\cite{droste2024explanations,unterbusch2023explanation,obaidi2025automatingexplanationneedmanagement}.

A key open question remains: \emph{To what extent can explanation needs be predicted from readily available app properties or metadata (e.g., app category, star ratings) rather than costly user studies or manual review analysis?}~\cite{Lu2017prado,brunotte2023privacy}. If such properties offer predictive value, developers and UX designers could proactively identify and address explanation needs, informing onboarding, help documentation, or targeted feature improvements and reducing the risk of unmet user expectations at scale. For example, certain app types (e.g., security apps, frequently updated tools) might consistently trigger explanation needs, allowing early prioritization for documentation or support resources.

In this work, we empirically investigate whether and how explanation needs, categorized from user reviews—correlate with or can be predicted from app properties. Using a large, expert-labeled dataset of 4,495 app reviews~\cite{goldstandard-explain-zenodo2024} enriched with metadata (app version, rating, age restriction, in-app purchases), we conduct correlation analyses and regression models to explore these relationships. Findings are validated with a recent, manually labeled sample of 495 reviews.

Our contributions are:
\begin{itemize}
    \item Extension and enrichment of an existing, labeled app review dataset~\cite{goldstandard-explain-zenodo2024} with relevant metadata for CrowdRE analysis.
    \item Large-scale correlation and regression analysis between app properties (e.g., genre, ratings, in-app purchases) and categorized explanation needs.
    \item Verification with an expert-labeled, recent review sample to assess generalizability.
    \item Reflection on the opportunities and limitations of inferring explanation needs from crowd-based app data.
\end{itemize}

The remainder of this paper is structured as follows: Section~\ref{sec:background} reviews background and related work; Section~\ref{sec:research} details the study design; Section~\ref{sec:results} presents results; and Section~\ref{sec:discussion} discusses implications before concluding in Section~\ref{sec:conclusion}.

\section{Background and Related Work}
\label{sec:background}

\subsection{Background}
\label{sec:erklearbarkeit}
\subsubsection{Explainability as a Software Quality Aspect}
Explainability describes how well users can understand different aspects of a system~\cite{chazette2021exploring,droste2024explanations}. While the concept stems from explainable AI, it is now recognized as relevant across diverse software domains~\cite{adadi2018peeking,kohl2019explainability,jongeling2024towards}. Recent studies have addressed explanation needs in areas such as privacy~\cite{brunotte2023context}, user onboarding~\cite{deters2023ondemand}, and workflows~\cite{deters2024UXandExplainability}.

However, not every explanation is helpful: inappropriate explanations can lead to cognitive overload or frustration~\cite{nunes2017systematic,chazette2020explainability}. Developers therefore need to understand which types of explanations are actually required.

\subsubsection{Explanation Needs in User Feedback}
One approach to uncovering user-facing explanation needs is the analysis of app reviews. Explanation needs may be explicit (e.g., direct questions) or implicit (e.g., confusion or missing information)~\cite{droste2024explanations}. While reviews are informal and not always focused on explanation, they offer scalable access to user perspectives despite inherent limitations (see Section~\ref{sec:discussion}).

Droste et al.~\cite{droste2024explanations} proposed a taxonomy of explanation needs across five system aspects, later expanded by Obaidi et al.~\cite{obaidi2025automatingexplanationneedmanagement} to include seven categories: \textit{Interaction}, \textit{System Behavior}, \textit{Domain Knowledge}, \textit{Security \& Privacy}, \textit{User Interface}, \textit{Business}, and \textit{Meta Information}.

Further studies explored user-specific factors. For example, Obaidi et al.~\cite{obaidi2025mood} examined the impact of mood on explanation needs and found only weak associations. Prior app knowledge may affect how users prefer explanations, but not whether explanations are needed~\cite{obaidi2025appKnowledge}.

\subsection{Related Work}
\subsubsection{Explainability Needs}
Various taxonomies for explanation needs have been proposed. Droste et al.~\cite{droste2024explanations} analyzed 315 user-submitted needs and found \textit{Interaction} to be the most frequent type, contrasting with AI literature, which often emphasizes \textit{System Behavior}. Sadeghi et al.~\cite{Sadeghi2021} focused on AI contexts and identified three explanation need categories related to disobedience, failure, and context-aware behavior.

Obaidi et al.~\cite{obaidi2025explainability} present a tool-supported approach for automatically deriving explainability requirements and explanations from user reviews. Their evaluation shows that while AI-generated explanations are often preferred for clarity and style, the relevance and correctness of AI-generated requirements lag behind manual ones, underscoring the ongoing need for human validation in explainability engineering.

Recent work by Obaidi et al.~\cite{obaidi2025elicit} empirically compared elicitation methods for explainability requirements, including focus groups, interviews, and surveys, and analyzed the impact of taxonomy usage. Their study found that interviews were most efficient, while surveys collected the most needs but with high redundancy. Importantly, combining elicitation methods and introducing taxonomies in a two-phase approach improved the number and diversity of explanation needs captured.

\subsubsection{App Review Analysis}
App reviews are widely used for analyzing user needs. Di Sorbo et al.~\cite{DiSorbo2020CorrelationApps} showed that bug reports negatively affect ratings, while Martin et al.~\cite{Martin2015appcorrelation} demonstrated that user requests are more robust indicators than app-level data. Biswas et al.~\cite{Biswas2021} related mHealth app quality to features and user sentiment using fuzzy logic.

In explainability research, Unterbusch et al.~\cite{unterbusch2023explanation} manually coded app reviews to derive explanation need categories, while Obaidi et al.~\cite{obaidi2025automatingexplanationneedmanagement} automated their detection and mapped needs to responsible teams. Pre-trained language models have also been used to classify reviews for RE tasks~\cite{Hadi2023}.

While our approach is aligned with these works, we aim to explore not just detection, but whether explanation needs can be inferred from app metadata alone—a question not yet systematically addressed in prior research.

\section{Study Design}
\label{sec:research}

\begin{figure}[htbp]
    \centering
    \includegraphics[width=1\linewidth]{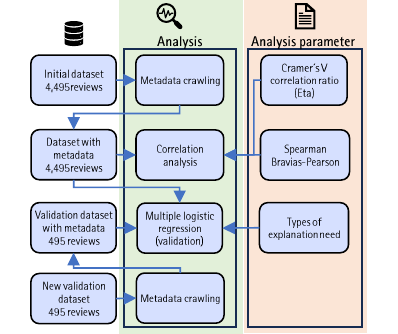}
    \caption{Implementation and evaluation of our study}
    \label{fig:online-studie}
\end{figure}

Our study follows a structured, two-step methodology. First, we crawled and integrated metadata with a gold standard dataset of 4,495 app reviews, enabling correlation analyses using methods such as Spearman, Pearson, Cramér’s V, and Eta. Second, we created a validation dataset of 495 reviews to evaluate the predictive power of app metadata on explanation needs via multiple logistic regression.

\subsection{Research Questions}

We address the following research questions:

\begin{itemize}
\item \textbf{RQ1: How do company-defined app properties relate to types of explanation needs?}  
This research question aims to determine whether objectively defined app properties set by companies influence explanation needs. If a strong correlation exists, explanation needs could be identified and addressed in advance.   

\item \textbf{RQ2: How do user feedback-derived app properties relate to types of explanation needs?}  
This question explores whether explanation needs evolve over the lifecycle of an app based on user feedback. 

\item \textbf{RQ3: How well can a combination of app properties predict explanation needs?}  
This research question seeks to determine whether explanation needs can be predicted using linear regression based on app properties. If feasible, this approach could enable the identification of explanation needs during the early stages of app development.   
\end{itemize}

\subsection{Data Preprocessing}

\begin{table*}[ht]
\small
\caption{Overview of hypotheses with corresponding variables.}
\label{tab:hypothesen-uebersicht}
\begin{tabularx}{\textwidth}{lXp{3.4cm}}
\toprule
\textbf{Hypothesis} & \textbf{Description (starts with: There is no relation between...)} & \textbf{Variables} \\ 
\midrule
H$1_0$ & explanation need (Yes/No) and app category & $E\textsubscript{needed}$, $C\textsubscript{category}$ \\ 
H$2_0$ (H$2.1_0$, H$2.2_0$) & explanation need types (explicit/implicit, categories) and app category & $E\textsubscript{type}$, $E\textsubscript{category}$, $C\textsubscript{category}$ \\ 
\midrule
H$3_0$ & explanation need (Yes/No) and app purchase type (free/paid) & $E\textsubscript{needed}$, $C\textsubscript{purch-type}$ \\ 
H$4_0$ (H$4.1_0$, H$4.2_0$) & explanation need types (explicit/implicit, categories) and app purchase type & $E\textsubscript{type}$, $E\textsubscript{category}$, $C\textsubscript{purch-type}$ \\ 
\midrule
H$5_0$ & explanation need (Yes/No) and app price & $E\textsubscript{needed}$, $C\textsubscript{purchase}$ \\ 
H$6_0$ (H$6.1_0$, H$6.2_0$) & explanation need types (explicit/implicit, categories) and app price & $E\textsubscript{type}$, $E\textsubscript{category}$, $C\textsubscript{purchase}$ \\ 
\midrule
H$7_0$ & explanation need (Yes/No) and in-game purchase price & $E\textsubscript{needed}$, $C\textsubscript{ingame}$ \\ 
H$8_0$ (H$8.1_0$, H$8.2_0$) & explanation need types (explicit/implicit, categories) and in-game purchase price & $E\textsubscript{type}$, $E\textsubscript{category}$, $C\textsubscript{ingame}$ \\ 
\midrule
H$9_0$ & explanation need (Yes/No) and app version & $E\textsubscript{needed}$, $C\textsubscript{version}$ \\ 
H$10_0$ (H$10.1_0$, H$10.2_0$) & explanation need types (explicit/implicit, categories) and app version & $E\textsubscript{type}$, $E\textsubscript{category}$, $C\textsubscript{version}$ \\ 
\midrule
H$11_0$ & explanation need (Yes/No) and app minimum age & $E\textsubscript{needed}$, $C\textsubscript{age}$ \\ 
H$12_0$ (H$12.1_0$, H$12.2_0$) & explanation need types (explicit/implicit, categories) and app minimum age & $E\textsubscript{type}$, $E\textsubscript{category}$, $C\textsubscript{age}$ \\ 
\midrule
H$13_0$ & explanation need (Yes/No) and app star rating & $E\textsubscript{needed}$, $U\textsubscript{rating}$ \\ 
H$14_0$ (H$14.1_0$, H$14.2_0$) & explanation need types (explicit/implicit, categories) and app star rating & $E\textsubscript{type}$, $E\textsubscript{category}$, $U\textsubscript{rating}$ \\ 
\midrule
H$15_0$ & explanation need (Yes/No) and number of app reviews & $E\textsubscript{needed}$, $U\textsubscript{reviews}$ \\ 
H$16_0$ (H$16.1_0$, H$16.2_0$) & explanation need types (explicit/implicit, categories) and number of app reviews & $E\textsubscript{type}$, $E\textsubscript{category}$, $U\textsubscript{reviews}$ \\ 
\midrule
H$17_0$ & explanation need (Yes/No) and number of app downloads & $E\textsubscript{needed}$, $U\textsubscript{downloads}$ \\ 
H$18_0$ (H$18.1_0$, H$18.2_0$) & explanation need types (explicit/implicit, categories) and app downloads number & $E\textsubscript{type}$, $E\textsubscript{category}$, $U\textsubscript{downloads}$ \\ 
\midrule
H$19_0$ - H$24_0$ & average explanation need and various app properties (category, price, in-game price, version, or minimum age) & $E\textsubscript{average}$, $C\textsubscript{category}$, $C\textsubscript{purchase}$, $C\textsubscript{ingame}$, $C\textsubscript{version}$, $C\textsubscript{age}$ \\ 
H$25_0$ - H$27_0$ & average explanation need and app star rating, reviews, or downloads & $E\textsubscript{average}$, $U\textsubscript{rating}$, $U\textsubscript{reviews}$, $U\textsubscript{downloads}$ \\ 
\bottomrule
\end{tabularx}
\end{table*}

We use the gold standard dataset from Obaidi~\cite{goldstandard-explain-zenodo2024}, which contains 4,495 app reviews categorized as implicit, explicit, or no need for explanation. These reviews are also mapped to the taxonomy by Droste et al.~\cite{droste2024explanations}. Metadata (e.g., category, age rating, rating, number of reviews, price, in-game price, version, installs) was collected via web crawling from the Apple App Store and Google Play Store.

\subsubsection{Validation Dataset}
\label{subsec:validieren-der-logistischen-regression}
For validation, we compiled 495 recent reviews from 10 apps (five from each store), with the last 50 reviews per app as of August 2024:
\begin{itemize}
    \item \textit{Apple App Store:} Subway Surfers, Telegram, DuckDuckGo Privacy Browser, Blitzer.de PLUS, Tinder
    \item \textit{Google Play Store:} Royal Match, Jolly Store, Things 3, AnkiDroid Flashcards, PeakFinder
\end{itemize}
Reviews were labeled by two experienced requirements engineers and were classified as explicit, implicit, or no explanation need, and assigned to taxonomy categories~\cite{droste2024explanations,Droste2025REJexpl}. Out of 495 reviews, 44 have explicit needs and 22 have implicit needs. The other 429 reviews do not contain explanation need. The distribution of the types of explanation needs are in Table \ref{tab:needsacross}. Discrepancies were resolved by consensus. Inter-rater agreement measured by Cohen’s Kappa was $\kappa = 0.89$ for explanation type and $\kappa = 0.88$ for taxonomy categories, indicating near-perfect agreement~\cite{cohen1960agreement,landis1977measurement}.

\begin{table}[t]
    \centering
    \caption{Distribution of explanation needs across 66 reviews (totaling 69 mentions); in three reviews two needs were coded.}
    \label{tab:needsacross}
    \small
    \setlength{\tabcolsep}{10pt}
    \begin{tabularx}{\columnwidth}{Xrr}
    \toprule
    \textbf{Type of explanation need}        & \textbf{\#}  & \textbf{\%}     \\ \midrule
    All needs combined           & 69          & 100.0\%         \\ \hline
    Interaction         & 14  & 20.6\% \\
    System Behavior     & 32  & 45.6\% \\
    User Interface      & 4  & 5.9\% \\ 
    Business            & 10  & 14.7\% \\
    Security \& Privacy & 4  & 5.9\%  \\
    Meta Information    & 2   & 2.9\%  \\ 
    Domain Knowledge    & 3   & 4.4\%  \\ \bottomrule
    \end{tabularx}
\end{table}

\subsection{Data Analysis}

\subsubsection{Variables}

Based on the annotated reviews and app metadata, we derived the variables listed in Table~\ref{tab:variables-overview}.

\begin{table}[ht]
    \centering
    \setlength{\tabcolsep}{4pt}
    \caption{Overview of variables, scales, and descriptions.}
    \label{tab:variables-overview}
    \small
    \begin{tabularx}{\columnwidth}{lllX}
        \toprule
        \textbf{Variable}   & \textbf{Scale} & \textbf{Range} & \textbf{Description} \\ 
        \midrule
        $E\textsubscript{needed}$ & Nominal & \{Yes; No\} & Explanation required \\ 
        $E\textsubscript{type}$ & Nominal & \{Explicit; Implicit\} & Type of need \\ 
        $E\textsubscript{category}$ & Nominal & \{Category\} & Explanation category \\ 
        $E\textsubscript{average}$ & Metric & $\mathbb{Q}_0^+$ & Avg. count of needs \\ \midrule
        $C\textsubscript{category}$ & Nominal & \{Category\} & App category \\ 
        $C\textsubscript{purch-type}$ & Nominal & \{Yes; No\} & Free or paid \\ 
        $C\textsubscript{purchase}$ & Metric & $\mathbb{Q}_0^+$ & App cost \\ 
        $C\textsubscript{version}$ & Ordinal & $\mathbb{Q}_0^+$ & App version \\ 
        $C\textsubscript{age}$ & Ordinal & \{4; 9; 12; 17\} & Age restriction \\ 
        $C\textsubscript{ingame}$ & Nominal & \{Yes; No\} & In-game purchases \\ \midrule
        $U\textsubscript{rating}$ & Ordinal & \{1, 2, 3, 4, 5\} & Star rating \\ 
        $U\textsubscript{reviews}$ & Metric & $n \in \mathbb{N}_0$ & No. of reviews \\ 
        $U\textsubscript{downloads}$ & Ordinal & $n \in \mathbb{N}_0$ & No. of downloads \\ 
        \bottomrule
    \end{tabularx}
\end{table}

Variables starting with $E$ denote explanation needs, $U$ for user feedback-derived properties, and $C$ for company-defined properties.

\subsubsection{Hypotheses Testing}
\label{subsec:hypothesen}

To systematically analyze relationships, we tested the hypotheses in Table~\ref{tab:hypothesen-uebersicht}. The hypotheses reflect typical app properties discussed in prior work~\cite{Martin2015appcorrelation,DiSorbo2020CorrelationApps,Biswas2021}. We applied Bonferroni correction to control for multiple testing~\cite{weisstein2004bonferroni}. 



\subsubsection{Correlation Analysis}
\label{sec:korrelationsanalyse}

After preparing the gold standard dataset, correlation analyses were conducted based on the null hypotheses. To perform these analyses, the following methods were used: Cramer's V~\cite{cramer1946mathematical}, the Eta coefficient~\cite{pearson1905fehlergesetz}, Spearman's rank correlation~\cite{spearman1904association}, and the Bravais-Pearson correlation~\cite{pearson1895regression}. The choice of correlation analysis method depended on the measurement scale of the variables, which were categorized accordingly (see Table~\ref{tab:variables-overview}). 

Cramer's V~\cite{cramer1946mathematical} and the Eta coefficient (correlation ratio)~\cite{pearson1905fehlergesetz} were employed to interpret the strength of the correlation for categorical and nominal data, while Spearman's rank correlation~\cite{spearman1904association} and the Bravais-Pearson correlation~\cite{pearson1895regression} were used for ordinal and metric data with a linear relationship. For Spearman's rank correlation, Cohen's effect size~\cite{Cohen1988} was applied to assess the magnitude of the relationship, while for the $\chi^2$ test, Cramer's V~\cite{cramer1946mathematical} value was calculated to evaluate the strength of association. The choice of method ensured that the analyses were suitable for the respective variable scales and relationships.

\section{Results}
\label{sec:results}

We present the results of the correlation analyses and regression models. All data is available at \href{https://doi.org/10.5281/zenodo.15851593}{Zenodo}~\cite{obaidi2025DatasetAppFeaturesExplainNeeds}.

\subsection{Correlation Analysis}
\label{sec:ergebnisse-der-korrelationsanalyse}

Table~\ref{tab:correlation_results} summarizes the key findings after Bonferroni correction. Hypotheses H1$_0$--H4$_0$, H6$_0$, H7$_0$--H14$_0$, H19$_0$, H20$_0$, H23$_0$--H24$_0$, and H26$_0$ were rejected, indicating statistically significant (but mostly weak or moderate) effects.

\begin{table}[ht]
    \centering
    \caption{Results of the correlation analysis.}
    \label{tab:correlation_results}
    \scriptsize
    \begin{tabularx}{\columnwidth}{X r r l l l}
        \toprule
        \textbf{H$_0$} & \textbf{Result} & \textbf{p-value} & \textbf{Method} & \textbf{Interpret.} & \textbf{Decision} \\
        \midrule
        H1$_0$ & 0.22 & $<$0.0001 & Cramér's V & Weak & Reject \\
        \midrule
        H2.1$_0$ & 0.11 & 0.2140 & Cramér's V & No & Do not reject \\
        H2.2$_0$ & 0.20 & $<$0.0001 & Cramér's V & Weak & Reject \\
        \midrule
        H3$_0$ & 0.12 & $<$0.0001 & Cramér's V & Weak & Reject \\
        \midrule
        H4.1$_0$ & 0.04 & 0.0410 & Cramér's V & No & Reject \\
        H4.2$_0$ & 0.17 & $<$0.0001 & Cramér's V & Weak & Reject \\
        \midrule
        H5$_0$ & 0.03 & 0.1807 & Eta & No & Do not reject \\
        \midrule
        H6.1$_0$ & 0.00 & 0.2622 & Eta & No & Do not reject \\
        H6.2$_0$ & 0.02 & $<$0.0001 & Eta & Weak & Reject \\
        \midrule
        H7$_0$ & 0.08 & 0.0036 & Cramér's V & No & Reject \\
        \midrule
        H8.1$_0$ & 0.14 & 0.0028 & Cramér's V & Weak & Reject \\
        H8.2$_0$ & 0.20 & $<$0.0001 & Cramér's V & Weak & Reject \\
        \midrule
        H9$_0$ & 0.39 & $<$0.0001 & Cramér's V & Moderate & Reject \\
        \midrule
        H10.1$_0$ & 0.26 & 0.0002 & Cramér's V & Moderate & Reject \\
        H10.2$_0$ & 0.34 & $<$0.0001 & Cramér's V & Moderate & Reject \\
        \midrule
        H11$_0$ & 0.13 & $<$0.0001 & Cramér's V & Weak & Reject \\
        \midrule
        H12.1$_0$ & 0.06 & 0.0410 & Cramér's V & No & Reject \\
        H12.2$_0$ & 0.09 & 0.0002 & Cramér's V & Weak & Reject \\
        \midrule
        H13$_0$ & 0.40 & $<$0.0001 & Cramér's V & Moderate & Reject \\
        \midrule
        H14.1$_0$ & 0.27 & 0.0019 & Cramér's V & Moderate & Reject \\
        H14.2$_0$ & 0.36 & $<$0.0001 & Cramér's V & Moderate & Reject \\
        \midrule
        H15$_0$ & 0.16 & 0.3997 & Eta & Weak & Do not reject \\
        \midrule
        H16.1$_0$ & 0.01 & 0.0007 & Eta & Weak & Reject \\
        H16.2$_0$ & 0.01 & $<$0.0001 & Eta & Weak & Reject \\
        \midrule
        H17$_0$ & 0.02 & 0.1471 & Eta & No & Do not reject \\
        \midrule
        H18.1$_0$ & 0.01 & 0.0903 & Eta & No & Do not reject \\
        H18.2$_0$ & 0.01 & 0.4815 & Eta & No & Do not reject \\
        \midrule
        H19$_0$ & -0.20 & 0.0249 & Spearman & Weak & Reject \\
        \midrule
        H20$_0$ & -0.22 & 0.0129 & Spearman & Weak & Reject \\
        \midrule
        H21$_0$ & -0.09 & 0.3353 & Pearson & No & Do not reject \\
        \midrule
        H22$_0$ & -0.23 & 0.0735 & Spearman & Weak & Do not reject \\
        \midrule
        H23$_0$ & 0.19 & 0.0398 & Spearman & Weak & Reject \\
        \midrule
        H24$_0$ & 0.26 & 0.0040 & Spearman & Weak & Reject \\
        \midrule
        H25$_0$ & 0.02 & 0.8427 & Spearman & No & Do not reject \\
        \midrule
        H26$_0$ & 0.38 & $<$0.0001 & Pearson & Moderate & Reject \\
        \midrule
        H27$_0$ & -0.17 & 0.1820 & Pearson & Weak & Do not reject \\
        \bottomrule
    \end{tabularx}
\end{table}

The correlation analysis examined potential relationships between explanation needs and various app properties. Several hypotheses could not be rejected, indicating that for many app properties, no significant correlation with explanation needs was found in our dataset.

Where significant results were observed, effect sizes were consistently weak to moderate. This supports prior findings that explanation needs are shaped by diverse and contextual factors~\cite{droste2024explanations,obaidi2025mood}.

For \textbf{company-defined properties} (RQ1), no significant correlation was found for app price, in-game purchases, or minimum age (H5$_0$, H6.1$_0$--H6.2$_0$, H7$_0$, H11$_0$, H12.1$_0$--H12.2$_0$, H19$_0$, H20$_0$, H21$_0$--H22$_0$, H25$_0$, H27$_0$). Thus, factors like free/paid status, price level, or age rating do not appear to strongly influence explanation needs.

For \textbf{feedback-derived properties} (RQ2), the number of downloads had no significant effect (H15$_0$, H16.2$_0$, H17$_0$, H18.1$_0$--H18.2$_0$, H23$_0$, H24$_0$), while the number of reviews showed weak correlations (H16.1$_0$), suggesting only a minor influence.

Among the significant findings, weak correlations were found for several variables: for company-defined properties, these included app category (H1$_0$, H2.2$_0$), purchase type (H3$_0$, H4.1$_0$, H4.2$_0$), and partially app version (H6.1$_0$, H8.1$_0$). Feedback-derived properties with weak effects included number of reviews and star rating (H13$_0$, H26$_0$).

Moderate correlations were observed only for app version (H9$_0$, H10.1$_0$, H10.2$_0$) and, among feedback-derived properties, for star ratings (H14.1$_0$, H14.2$_0$).


\vspace{3pt}
\setlength{\shadowsize}{2pt}
\noindent
\shadowbox{
\begin{minipage}{0.94\columnwidth}
    \textbf{Finding:} Correlation analysis shows that only a few app properties—mainly app version, star rating, and review count—exhibit weak-to-moderate relationships with explanation needs. These factors offer tentative signals, but are insufficient as sole indicators for design decisions.
\end{minipage}
}

\subsection{Linear Regression}
\label{subsec:results-linear-regression}

Table~\ref{tab:regression_results} summarizes the regression results for both the complete dataset and the individual app store subsets. 

\begin{table}[ht]
\centering
\scriptsize
\caption{Results of the Linear Regressions}
\label{tab:regression_results}
\begin{tabularx}{\columnwidth}{Xlrrr}
\toprule
\textbf{Predicted} & 
\textbf{Dataset} & 
\textbf{Score} & 
\textbf{RMSE} & 
\boldmath$r^2$ \\
\midrule
$E\textsubscript{type}$ & Both Stores & 0.52 & 0.95 & -0.646 \\
$E\textsubscript{type}$ & Playstore & 0.57 & 0.90 & -0.625 \\
$E\textsubscript{type}$ & Appstore & 0.51 & 0.93 & -0.536 \\
\midrule
$E\textsubscript{cat}$ (Business) & Both Stores & 0.83 & 0.40 & 0.703 \\
$E\textsubscript{cat}$ (Domain Knowledge) & Both Stores & 0.82 & 0.40 & 0.702 \\
$E\textsubscript{cat}$ (Interaction) & Both Stores & 0.78 & 0.45 & 0.647 \\
$E\textsubscript{cat}$ (Meta information) & Both Stores & 0.79 & 0.43 & 0.678 \\
$E\textsubscript{cat}$ (Security \& Privacy) & Both Stores & 0.83 & 0.40 & 0.701 \\
$E\textsubscript{cat}$ (System behavior) & Both Stores & 0.81 & 0.42 & 0.678 \\
$E\textsubscript{cat}$ (User Interface) & Both Stores & 0.77 & 0.46 & 0.642 \\
\bottomrule
\end{tabularx}
\end{table}

The results show that linear regression models based solely on app properties yielded poor predictive power for the general classification of explanation needs (explicit, implicit, none), with negative \( r^2 \) values across all datasets. 

For explanation need categories, moderate \( r^2 \) values were observed in \textit{Business} (0.703), \textit{Domain Knowledge} (0.702), and \textit{Security \& Privacy} (0.701), while other categories remained lower (\( r^2 = 0.642 - 0.678 \)). 


\vspace{3pt}
\setlength{\shadowsize}{2pt}
\noindent
\shadowbox{
\begin{minipage}{0.94\columnwidth}
    \textbf{Finding:} Linear regression shows that app properties alone are insufficient to reliably predict explanation needs. Most categories remain difficult to predict and require additional contextual information.
\end{minipage}
}

\subsubsection{Validation}

Table~\ref{tab:regression_validierung} summarizes the validation results for the linear regression models. Note that perfect scores (\( \text{Score} = 1 \), RMSE = 0, \( r^2 = 1 \)) occur for categories not present in the validation set and are not meaningful. Higher scores may also reflect the smaller sample size in validation.

\begin{table}[ht]
\centering
\scriptsize
\caption{Validation Results of the Linear Regressions}
\label{tab:regression_validierung}
\begin{tabularx}{\columnwidth}{Xlrrr}
\toprule
\textbf{Predicted} & 
\textbf{Dataset} & 
\textbf{Score} & 
\textbf{RMSE} & 
\boldmath$r^2$ \\
\midrule
$E\textsubscript{type}$ & Both Stores & 0.86 & 0.52 & -0.137 \\
$E\textsubscript{type}$ & Playstore & 0.87 & 0.54 & -0.133 \\
$E\textsubscript{type}$ & Appstore & 0.86 & 0.50 & -0.141 \\
\midrule
$E\textsubscript{cat}$ (Business)  & Both Stores & 0.98 & 0.14 & -0.021 \\
$E\textsubscript{cat}$ (Domain Knowledge)  & Both Stores & 0.99 & 0.08 & -0.006 \\
$E\textsubscript{cat}$ (Interaction)  & Both Stores & 0.98 & 0.16 & 0.118 \\
$E\textsubscript{cat}$ (Meta information)  & Both Stores & 1.00 & 0.06 & -0.004 \\
$E\textsubscript{cat}$ (System behavior)  & Both Stores & 0.94 & 0.25 & -0.067 \\
$E\textsubscript{cat}$ (Security \& Privacy)  & Both Stores & 0.99 & 0.09 & -0.008 \\
$E\textsubscript{cat}$ (User Interface)  & Both Stores & 0.99 & 0.09 & -0.008 \\
\bottomrule
\end{tabularx}
\end{table}

The validation set confirms that linear regression models based on app properties perform poorly in practice. Classification of explanation needs (explicit, implicit, none) again resulted in negative \( r^2 \) values, indicating that the models do not generalize beyond the original dataset.

Among the explanation need categories, \textit{Interaction} achieved the highest (but still low) \( r^2 \) of 0.118. Other categories showed negligible or negative \( r^2 \) values, with \textit{Business} (\(-0.021\)), \textit{Domain Knowledge} (\(-0.006\)), and \textit{Meta Information} (\(-0.004\)) at best.


\vspace{3pt}
\setlength{\shadowsize}{2pt}
\noindent
\shadowbox{
\begin{minipage}{0.94\columnwidth}
	\textbf{Finding:} Validation confirms that linear regression on app properties does not generalize well for explanation need prediction. Most categories show only weak or negligible predictive power, emphasizing the need for additional features or alternative methods.
\end{minipage}
}

\section{Discussion}
\label{sec:discussion}

In the following, we answer the research questions, present threats to validity, and interpret the results.

\subsection{Answering the Research Questions}
\label{sec:beantworten-der-forschungsfragen}

\textbf{RQ1: How do company-defined app properties relate to types of explanation needs?}\\
Company-defined app properties (e.g., app category, price, version) show mostly weak correlations with explanation needs. Only app version exhibited moderate correlations (H9$_0$, H10.1$_0$–H10.2$_0$), suggesting that newer or frequently updated apps may trigger different explanation needs. For most properties, however, no strong relationship was identified.

\textbf{RQ2: How do user feedback-derived app properties relate to types of explanation needs?}\\
User feedback-derived properties such as review volume and star ratings showed limited, mostly weak-to-moderate correlations with explanation needs (e.g., H14.1$_0$–H14.2$_0$, H16.1$_0$, H16.2$_0$, H26$_0$). No significant correlation was observed for other feedback-derived properties, indicating that user feedback alone does not strongly predict explanation needs.

\textbf{RQ3: How well can a combination of app properties predict explanation needs?}\\
Combinations of app properties do not yield sufficient predictive power: Linear regression models resulted in negative or low \( r^2 \) values for all configurations. Thus, explanation needs cannot be reliably inferred from app metadata alone.

\subsection{Interpretation}
\label{sec:interpretation}

Our findings reveal that the relationship between app properties and explanation needs is more nuanced and context-dependent than initially expected. The correlation analyses show that most company-defined properties (such as app category, price, or minimum age) and feedback-derived properties (such as download numbers) are only weakly or inconsistently associated with explanation needs. The few notable exceptions—app version, star ratings, and review volume—exhibit moderate correlations with certain explanation need types, suggesting that frequent updates or changes in user sentiment can occasionally trigger more pronounced information needs. This pattern aligns with prior work indicating that explanation needs are shaped by a complex interplay of technical, contextual, and user-specific factors~\cite{droste2024explanations,obaidi2025mood}.

The linear regression models reinforce this complexity: app metadata alone cannot reliably predict explanation needs, and only some categories (such as \textit{Business} or \textit{Domain Knowledge}) show moderate predictive power. Most explanation need types—particularly those that are highly contextual or less frequently occurring—remain difficult to infer based on surface-level features alone. These findings are robust across both the gold standard and validation datasets, underlining the limitations of automated requirements identification that relies solely on app properties.

While our results may seem unsurprising in that metadata lacks the nuance to capture user-specific informational needs, this negative finding offers an important theoretical insight: explanation needs often reflect latent, context-rich user expectations that are not encoded in static app properties. This emphasizes the inherent limits of surface-level prediction and motivates a shift toward richer, multi-modal data sources. For example, incorporating natural language processing (NLP) on review content, mining behavioral telemetry, or leveraging usage traces could reveal deeper signals of explanation needs that app properties alone miss.

\textit{Implications for stakeholders:} 
For software developers and requirements engineers, these results suggest that easily accessible app metadata—like version changes or fluctuating star ratings—may serve as preliminary signals to flag potential explanation needs. For example, after a major update, teams might proactively review user feedback for new explanation needs or prepare additional onboarding materials. However, such metadata should be viewed as supplementary: it cannot replace direct user engagement, usability studies, or qualitative review analysis. Effective requirements engineering still relies on active dialogue with users and systematic review mining to capture the diversity and dynamics of explanation needs.

For CrowdRE practitioners, the findings highlight that large-scale mining of app properties can help in the early identification of areas worth investigating further, but also underscore that actionable requirements must be grounded in more detailed and context-aware user analysis. In practice, this means metadata analysis should be integrated with review mining, sentiment analysis, or real-time behavioral data to triangulate emerging needs and support prioritization. The limited predictive power found here also suggests that CrowdRE workflows could benefit from automation that combines structured metadata with unstructured user feedback, such as using embeddings from review texts or clustering behavioral patterns.

For researchers, these findings point to the value of exploring not just more metadata, but qualitatively different sources and more advanced analytic techniques. Future work should pivot toward hybrid models and methods—leveraging machine learning on review content or behavioral telemetry—to improve the detection and operationalization of explanation needs, and to ultimately bridge the gap between static app properties and user-centered requirements.

\subsection{Threats to Validity}
\label{sec:validiteat}

We discuss threats to validity following Wohlin~\cite{wohlin2012experimentation}, focusing on construct, internal, conclusion, and external validity.

\textbf{Construct Validity.}
While the gold standard dataset was systematically labeled and reviewed, labeling errors or misclassifications may persist, especially given the subjective nature of explanation needs. Our taxonomy choice~\cite{droste2024explanations} also constrains possible results; using alternative categorization schemes could yield different findings. The validation dataset, though labeled by experienced raters with high agreement ($\kappa = 0.89$), remains limited in size and diversity, which may affect the detection of rare explanation need types. The application of the Bonferroni correction reduced Type I error but increased the risk of Type II errors, potentially obscuring weaker relationships. Not all app categories or property combinations are represented, limiting the diversity of explanation needs captured.

\textbf{Internal Validity.}
Errors may have occurred in matching metadata to reviews or during data preprocessing. While two researchers interpreted results and multiple rounds of annotation were conducted, involving more independent analysts could further reduce interpretation bias and improve reliability.

\textbf{Conclusion Validity.}
The relatively small validation dataset and the underrepresentation of certain explanation need categories limit the robustness of statistical conclusions, especially for rare cases like implicit needs. Similar gaps in the gold standard dataset reduce the generalizability of findings. Weak-to-moderate correlations suggest unmeasured influencing factors; future work should further connect empirical analysis to existing theory for deeper interpretation.

\textbf{External Validity.}
Our datasets cover only a subset of apps and explanation need categories, limiting the generalizability to other domains or software ecosystems. The taxonomy used, while systematically developed, may not capture all relevant explanation needs; different taxonomies or more diverse datasets could yield other insights. Broader studies are needed to validate these results in other CrowdRE or app contexts.

\subsection{Future Work}
\label{sec:ausblick}

Future research should address several key limitations of this study. Most urgently, expanding the dataset—especially to better represent rarely occurring explanation need categories and subcategories—would improve both the robustness and generalizability of the results. This is particularly relevant for underrepresented types such as implicit needs, where prediction and classification remain unreliable.

Beyond dataset size, incorporating additional app properties such as update frequency, feature complexity, or user demographics could provide richer signals for predicting explanation needs. However, our findings also indicate that many commonly available metadata features are insufficient, and thus future studies should prioritize richer, multi-modal data, such as NLP analysis of review text, behavioral telemetry, or in-app usage analytics. These approaches may uncover deeper and more actionable signals, helping to operationalize explainability requirements in practice.

On the methodological side, future work should benchmark metadata-based prediction against more advanced models (e.g., Random Forests, Gradient Boosting, neural networks, or LLM-based approaches) and compare findings with related work (e.g.,~\cite{Hadi2023, Sadeghi2021}) to better position results in the literature. Hybrid models that combine metadata and textual review analysis, such as embeddings from review text, may further improve predictive accuracy. Exploratory analyses like clustering or dimensionality reduction could also help uncover latent explanation need profiles in app populations.

Finally, future research should translate empirical findings into practical guidelines for requirements engineering, CrowdRE, and app development. This may include strategies for integrating explanation need analysis into early requirements elicitation, onboarding, or user guidance—supporting more explainable and user-centric software.

\section{Conclusion}
\label{sec:conclusion}

This study examined whether app properties, such as genre, platform, ratings, and review counts, can predict users' explanation needs in app reviews. Analyzing a gold standard dataset of 4,495 annotated reviews with app metadata and a validation sample, we found that most app properties exhibit only weak or inconsistent correlations with explanation needs; only app version, star ratings, and review counts showed moderate associations. Linear regression models further demonstrated that app metadata alone does not reliably predict explanation needs, with most categories yielding low or negative \( r^2 \) values. Only categories like \textit{Business} and \textit{Security \& Privacy} showed slightly higher predictability, but overall, context-specific information is essential. For practitioners, app properties may serve as preliminary signals to inform priorities for onboarding, help, or documentation, but should not replace direct user engagement. Developers should supplement metadata insights with active user feedback to address diverse explanation needs. For researchers, these findings highlight the limitations of surface-level app data for predicting user expectations and point to the need for richer datasets, additional features, and advanced models. Future work should integrate contextual and behavioral factors, leverage modern machine learning approaches, and provide practical guidelines for explainability-driven requirements engineering.

\section*{Acknowledgment}
This work was funded by the Deutsche Forschungsgemeinschaft (DFG, German Research Foundation) under Grant No.: 470146331, project softXplain (2022-2025).

\bibliographystyle{IEEEtran}
\bibliography{references.bib}

\begin{thebibliography}{10}
\providecommand{\url}[1]{#1}
\csname url@samestyle\endcsname
\providecommand{\newblock}{\relax}
\providecommand{\bibinfo}[2]{#2}
\providecommand{\BIBentrySTDinterwordspacing}{\spaceskip=0pt\relax}
\providecommand{\BIBentryALTinterwordstretchfactor}{4}
\providecommand{\BIBentryALTinterwordspacing}{\spaceskip=\fontdimen2\font plus
\BIBentryALTinterwordstretchfactor\fontdimen3\font minus
  \fontdimen4\font\relax}
\providecommand{\BIBforeignlanguage}[2]{{%
\expandafter\ifx\csname l@#1\endcsname\relax
\typeout{** WARNING: IEEEtran.bst: No hyphenation pattern has been}%
\typeout{** loaded for the language `#1'. Using the pattern for}%
\typeout{** the default language instead.}%
\else
\language=\csname l@#1\endcsname
\fi
#2}}
\providecommand{\BIBdecl}{\relax}
\BIBdecl

\bibitem{adadi2018peeking}
A.~Adadi and M.~Berrada, ``Peeking inside the black-box: a survey on
  explainable artificial intelligence (xai),'' \emph{IEEE access}, vol.~6, pp.
  52\,138--52\,160, 2018.

\bibitem{Antinyan2020complex}
V.~Antinyan, ``Revealing the complexity of automotive software,'' in
  \emph{ESEC/FSE'20}.\hskip 1em plus 0.5em minus 0.4em\relax Association for
  Computing Machinery, 2020.

\bibitem{levy2021understanding}
O.~Levy and D.~Feitelson, ``Understanding large-scale software systems –
  structure and flows,'' \emph{Empirical Software Engineering}, vol.~26, no.~1,
  2021.

\bibitem{kohl2019explainability}
M.~A. Köhl, K.~Baum, M.~Langer, D.~Oster, T.~Speith, and D.~Bohlender,
  ``Explainability as a non-functional requirement,'' in \emph{RE'19}, 2019.

\bibitem{chazette2021exploring}
L.~Chazette, W.~Brunotte, and T.~Speith, ``Exploring explainability: a
  definition, a model, and a knowledge catalogue,'' in \emph{RE}.\hskip 1em
  plus 0.5em minus 0.4em\relax IEEE, 2021.

\bibitem{Deters2025quality}
H.~Deters, J.~Droste, M.~Obaidi, and K.~Schneider, ``Exploring the means to
  measure explainability: Metrics, heuristics and questionnaires,''
  \emph{Information and Software Technology}, vol. 181, p. 107682, 2025.

\bibitem{deters2024qualitymodel}
------, ``How explainable is your system? towards a quality model
  for explainability,'' in \emph{Requirements Engineering: Foundation for
  Software Quality}, D.~Mendez and A.~Moreira, Eds.\hskip 1em plus 0.5em minus
  0.4em\relax Springer Nature Switzerland, 2024, pp. 3--19.

\bibitem{deters2025identifying}
H.~Deters, L.~Reinhardt, J.~Droste, M.~Obaidi, and K.~Schneider, ``Identifying
  explanation needs: Towards a catalog of user-based indicators,'' in
  \emph{2025 IEEE 33rd International Requirements Engineering Conference (RE)},
  Valencia, Spain, Sep. 2025.

\bibitem{Biswas2021}
M.~Biswas, M.~H. Tania, M.~S. Kaiser, R.~Kabir, M.~Mahmud, and A.~A. Kemal,
  ``Accu3rate: A mobile health application rating scale based on user
  reviews,'' \emph{PloS one}, vol.~16, no.~12, p. e0258050, 2021.

\bibitem{brunotte2023privacy}
W.~Brunotte, A.~Specht, L.~Chazette, and K.~Schneider, ``Privacy
  explanations--a means to end-user trust,'' \emph{JSS}, vol. 195, 2023.

\bibitem{brunotte_quo_2022}
W.~Brunotte, L.~Chazette, V.~Kl{\"o}s, and T.~Speith, ``Quo vadis,
  explainability? -- a research roadmap for explainability engineering,'' in
  \emph{Requirements Engineering: Foundation for Software Quality}.\hskip 1em
  plus 0.5em minus 0.4em\relax Cham: Springer International Publishing, 2022,
  pp. 26--32.

\bibitem{Frattini22}
J.~Frattini, L.~Montgomery, J.~Fischbach, M.~Unterkalmsteiner, D.~Mendez, and
  D.~Fucci, ``A live extensible ontology of quality factors for textual
  requirements,'' in \emph{2022 IEEE 30th International Requirements
  Engineering Conference (RE)}, 2022, pp. 274--280.

\bibitem{chazette2022can}
L.~Chazette, J.~Kl{\"u}nder, M.~Balci, and K.~Schneider, ``How can we develop
  explainable systems? insights from a literature review and an interview
  study,'' in \emph{Proceedings of the International Conference on Software and
  System Processes and International Conference on Global Software
  Engineering}, 2022, pp. 1--12.

\bibitem{obaidi2025appKnowledge}
M.~Obaidi, J.~Fischbach, M.~Herrmann, H.~Deters, J.~Droste, J.~Kl{\"u}nder, and
  K.~Schneider, ``How does users' app knowledge influence the preferred level
  of detail and format of software explanations?'' in \emph{Requirements
  Engineering: Foundation for Software Quality: 31st International Working
  Conference}.\hskip 1em plus 0.5em minus 0.4em\relax Springer Nature
  Switzerland, 2025.

\bibitem{anders2022userfeedback}
M.~Anders, M.~Obaidi, B.~Paech, and K.~Schneider, ``A study on the mental
  models of users concerning existing software,'' in \emph{Requirements
  Engineering: Foundation for Software Quality}, V.~Gervasi and A.~Vogelsang,
  Eds.\hskip 1em plus 0.5em minus 0.4em\relax Cham: Springer International
  Publishing, 2022, pp. 235--250.

\bibitem{anders2023userfeedback}
M.~Anders, M.~Obaidi, A.~Specht, and B.~Paech, ``What can be concluded from
  user feedback? - an empirical study,'' in \emph{2023 IEEE 31st International
  Requirements Engineering Conference Workshops (REW)}, 2023, pp. 122--128.

\bibitem{obaidi2025elicit}
M.~Obaidi, J.~Droste, H.~Deters, M.~Herrmann, R.~Ochsner, J.~Klünder, and
  K.~Schneider, ``How to elicit explainability requirements? a comparison of
  interviews, focus groups, and surveys,'' in \emph{2025 IEEE 33rd
  International Requirements Engineering Conference (RE)}, Valencia, Spain,
  Sep. 2025.

\bibitem{unterbusch2023explanation}
M.~Unterbusch, M.~Sadeghi, J.~Fischbach, M.~Obaidi, and A.~Vogelsang,
  ``Explanation needs in app reviews: Taxonomy and automated detection,'' in
  \emph{2023 IEEE 31st International Requirements Engineering Conference
  Workshops (REW)}.\hskip 1em plus 0.5em minus 0.4em\relax IEEE, 2023.

\bibitem{obaidi2025automatingexplanationneedmanagement}
M.~Obaidi, N.~Voß, J.~Droste, H.~Deters, M.~Herrmann, J.~Fischbach, and
  K.~Schneider, ``Automating explanation need management in app reviews: A case
  study from the navigation app industry,'' in \emph{Proceedings of the 47th
  International Conference on Software Engineering: Software Engineering in
  Practice}, ser. ICSE-SEIP'25.\hskip 1em plus 0.5em minus 0.4em\relax New
  York, NY, USA: Association for Computing Machinery, 2025.

\bibitem{droste2023designing}
J.~Droste, H.~Deters, J.~Puglisi, and J.~Kl{\"u}nder, ``Designing end-user
  personas for explainability requirements using mixed methods research,'' in
  \emph{REW}.\hskip 1em plus 0.5em minus 0.4em\relax IEEE, 2023.

\bibitem{ramos2021modeling}
H.~Ramos, M.~Fonseca, and L.~Ponciano, ``Modeling and evaluating personas with
  software explainability requirements,'' in \emph{HCI-COLLAB'21}.\hskip 1em
  plus 0.5em minus 0.4em\relax Springer, 2021.

\bibitem{xu2019roleOfUserMood}
L.~Xu, X.~Zhou, and U.~Gadiraju, ``Revealing the role of user moods in
  struggling search tasks,'' in \emph{SIGIR'19}.\hskip 1em plus 0.5em minus
  0.4em\relax Association for Computing Machinery, 2019.

\bibitem{droste2024explanations}
J.~Droste, H.~Deters, M.~Obaidi, and K.~Schneider, ``Explanations in everyday
  software systems: Towards a taxonomy for explainability needs,'' in
  \emph{2024 IEEE 32nd International Requirements Engineering Conference (RE)},
  2024, pp. 55--66.

\bibitem{Droste2025REJexpl}
\BIBentryALTinterwordspacing
J.~Droste, H.~Deters, M.~Obaidi, J.~Klünder, and K.~Schneider, ``Framing what
  can be explained -- an operational taxonomy for explainability needs,''
  \emph{Requirements Engineering}, 2025. [Online]. Available:
  \url{https://doi.org/10.1007/s00766-025-00440-x}
\BIBentrySTDinterwordspacing

\bibitem{Lu2017prado}
X.~Lu, Z.~Chen, X.~Liu, H.~Li, T.~Xie, and Q.~Mei, ``Prado: Predicting app
  adoption by learning the correlation between developer-controllable
  properties and user behaviors,'' \emph{Proc. ACM Interact. Mob. Wearable
  Ubiquitous Technol.}, vol.~1, no.~3, sep 2017.

\bibitem{goldstandard-explain-zenodo2024}
\BIBentryALTinterwordspacing
M.~Obaidi, ``{Dataset: Gold standard dataset for explainability need detection
  in app reviews.}'' Sep. 2024. [Online]. Available:
  \url{https://doi.org/10.5281/zenodo.11522828}
\BIBentrySTDinterwordspacing

\bibitem{jongeling2024towards}
R.~Jongeling, ``Towards public understanding of software through modeling,'' in
  \emph{Proceedings of the ACM/IEEE 27th International Conference on Model
  Driven Engineering Languages and Systems}, 2024, pp. 665--669.

\bibitem{brunotte2023context}
W.~Brunotte, J.~Droste, and K.~Schneider, ``Context, content, consent-how to
  design user-centered privacy explanations (s).'' in \emph{SEKE}, 2023, pp.
  86--89.

\bibitem{deters2023ondemand}
H.~Deters, J.~Droste, M.~Fechner, and J.~Kl{\"u}nder, ``Explanations on
  demand-a technique for eliciting the actual need for explanations,'' in
  \emph{REW}.\hskip 1em plus 0.5em minus 0.4em\relax IEEE, 2023.

\bibitem{deters2024UXandExplainability}
H.~Deters, J.~Droste, A.~Hess, V.~Kl\"{o}s, K.~Schneider, T.~Speith, and
  A.~Vogelsang, ``The x factor: On the relationship between user experience and
  explainability,'' in \emph{NordiCHI'24}.\hskip 1em plus 0.5em minus
  0.4em\relax Association for Computing Machinery, 2024.

\bibitem{nunes2017systematic}
I.~Nunes and D.~Jannach, ``A systematic review and taxonomy of explanations in
  decision support and recommender systems,'' \emph{User Modeling and
  User-Adapted Interaction}, vol.~27, 2017.

\bibitem{chazette2020explainability}
L.~Chazette and K.~Schneider, ``Explainability as a non-functional requirement:
  challenges and recommendations,'' \emph{REJ}, vol.~25, no.~4, 2020.

\bibitem{obaidi2025mood}
M.~Obaidi, J.~Droste, H.~Deters, M.~Herrmann, J.~Kl{\"u}nder, and K.~Schneider,
  ``Do users' explainability needs in software change with mood?'' in
  \emph{Requirements Engineering: Foundation for Software Quality: 31st
  International Working Conference}.\hskip 1em plus 0.5em minus 0.4em\relax
  Springer Nature Switzerland, 2025.

\bibitem{Sadeghi2021}
M.~Sadeghi, V.~Kl{\"o}s, and A.~Vogelsang, ``Cases for explainable software
  systems: Characteristics and examples,'' in \emph{2021 IEEE 29th
  International Requirements Engineering Conference Workshops (REW)}.\hskip 1em
  plus 0.5em minus 0.4em\relax IEEE, 2021, pp. 181--187.

\bibitem{obaidi2025explainability}
M.~Obaidi, J.~Fischbach, J.~Droste, H.~Deters, M.~Herrmann, J.~Klünder,
  S.~Krätzig, H.~Villamizar, and K.~Schneider, ``Automatic generation of
  explainability requirements and software explanations from user reviews,'' in
  \emph{2025 IEEE 33rd International Requirements Engineering Conference
  Workshops (REW)}, 2025.

\bibitem{DiSorbo2020CorrelationApps}
A.~D. Sorbo, G.~Grano, C.~A. Visaggio, and S.~Panichella, ``Investigating the
  criticality of user-reported issues through their relations with app
  rating,'' \emph{Journal of Software: Evolution and Process}, 2020.

\bibitem{Martin2015appcorrelation}
W.~Martin, M.~Harman, Y.~Jia, F.~Sarro, and Y.~Zhang, ``The app sampling
  problem for app store mining,'' in \emph{2015 IEEE/ACM 12th Working
  Conference on Mining Software Repositories}, 2015, pp. 123--133.

\bibitem{Hadi2023}
M.~A. Hadi and F.~H. Fard, ``Evaluating pre-trained models for user feedback
  analysis in software engineering: a study on classification of app-reviews,''
  \emph{Empirical Software Engineering}, vol.~28, no.~4, p.~88, May 2023.

\bibitem{cohen1960agreement}
J.~Cohen, ``A coefficient of agreement for nominal scales,'' \emph{Educational
  and Psychological Measurement}, vol.~20, no.~1, 1960.

\bibitem{landis1977measurement}
J.~Landis and G.~Koch, ``The measurement of observer agreement for categorical
  data.'' \emph{Biometrics}, vol. 33 1, 1977.

\bibitem{weisstein2004bonferroni}
E.~W. Weisstein, ``Bonferroni correction,'' \emph{Wolfram Research, Inc.},
  2004, accessed on 03.27.2024, 03:50 AM.

\bibitem{cramer1946mathematical}
H.~Cramér, \emph{Mathematical Methods of Statistics}.\hskip 1em plus 0.5em
  minus 0.4em\relax Princeton, NJ: Princeton University Press, 1946.

\bibitem{pearson1905fehlergesetz}
K.~Pearson, ``Das fehlergesetz und seine verallgemeinerungen durch fechner und
  pearson: A rejoinder,'' \emph{Biometrika}, vol.~4, no. 1-2, pp. 169--212,
  June 1905.

\bibitem{spearman1904association}
C.~Spearman, ``The proof and measurement of association between two things,''
  \emph{The American Journal of Psychology}, vol.~15, no.~1, pp. 72--101, 1904.

\bibitem{pearson1895regression}
K.~Pearson, ``Note on regression and inheritance in the case of two parents,''
  \emph{Proceedings of the Royal Society of London}, vol.~58, pp. 240--242,
  1895.

\bibitem{Cohen1988}
J.~Cohen, \emph{Statistical Power Analysis for the Behavioral Sciences},
  2nd~ed.\hskip 1em plus 0.5em minus 0.4em\relax Routledge, 1988.

\bibitem{obaidi2025DatasetAppFeaturesExplainNeeds}
\BIBentryALTinterwordspacing
M.~Obaidi, Q.~Kushtrim, J.~Droste, H.~Deters, M.~Herrmann, J.~Klünder,
  E.~Schmid, and K.~Schneider, ``Dataset: From app features to explanation
  needs: Analyzing correlations and predictive potential,'' Jul. 2025.
  [Online]. Available: \url{https://doi.org/10.5281/zenodo.15851593}
\BIBentrySTDinterwordspacing

\bibitem{wohlin2012experimentation}
C.~Wohlin, P.~Runeson, M.~Höst, M.~C. Ohlsson, B.~Regnell, and A.~Wesslén,
  \emph{Experimentation in software engineering}.\hskip 1em plus 0.5em minus
  0.4em\relax Springer, 2012.

\end{thebibliography}

\end{document}